\documentclass[]{spie}  

\renewcommand{\baselinestretch}{1.65} 
\usepackage{subcaption}
\usepackage{setspace}

\usepackage{amsmath,amsfonts,amssymb}
\usepackage{graphicx}
\usepackage[colorlinks=true, allcolors=blue]{hyperref}
\usepackage[leftcaption]{sidecap}

\title{Investigation of large-scale extended Granger causality (lsXGC) on synthetic functional MRI data}

\author[a,b,c,d]{Axel Wismüller}
\author[a]{M.~Ali Vosoughi}
\author[a]{Adora DSouza}
\author[c]{Anas Z.~Abidin}

\affil[a]{Department of Electrical and Computer Engineering, University of Rochester, NY, USA}
\affil[b]{ Department of Imaging Sciences, University of Rochester, NY, USA}
\affil[c]{Department of Biomedical Engineering, University of Rochester, NY, USA}
\affil[d]{Faculty of Medicine and Institute of Clinical Radiology, Ludwig Maximilian University,
Munich, Germany}

\authorinfo{Further author information: (Send correspondence to Ali Vosoughi)\\Ali Vosoughi: E-mail: mvosough@ece.rochester.edu}

\begin{document} 
\maketitle

\begin{abstract}

It is a challenging research endeavor to infer causal relationships in multivariate observational time-series. Such data may be represented by graphs, where nodes represent time-series, and edges directed causal influence scores between them. If the number of nodes exceeds the number of temporal observations, conventional methods, such as standard Granger causality, are of limited value, because estimating free parameters of time-series predictors lead to underdetermined problems. A typical example for this situation is functional Magnetic Resonance Imaging (fMRI), where the number of nodal observations is large, usually ranging from $10^2$ to $10^5$ time-series, while the number of temporal observations is low, usually less than $10^3$. Hence, innovative approaches are required to address the challenges arising from such data sets. Recently, we have proposed the large-scale Extended Granger Causality (lsXGC) algorithm, which is based on augmenting a dimensionality-reduced representation of the system's state-space by supplementing data from the conditional source time-series taken from the original input space. Here, we apply lsXGC on synthetic fMRI data with known ground truth and compare its performance to state-of-the-art methods by leveraging the benefits of information-theoretic approaches. Our results suggest that the proposed lsXGC method significantly outperforms existing methods, both in diagnostic accuracy with Area Under the Receiver Operating Characteristic (AUROC = $0.849$ vs.~$[0.727, 0.762]$ for competing methods, $p<\!10^{-8}$), and computation time ($3.4$ sec vs.~[$9.7$, $4.8 \times 10^3$] sec for competing methods) benchmarks, demonstrating the potential of lsXGC for analyzing large-scale networks in neuroimaging studies of the human brain.

\end{abstract}


\section{INTRODUCTION} \label{sec:intro}  

For studying the underlying pathophysiology of neurological and psychiatric disease, functional connectivity among various cortical regions has been identified as an important research subject [\citeonline{4_norman2006beyond}]. By providing images with sufficiently high spatial resolution and the hemodynamic response dynamics over a temporal axis, fMRI studies have demonstrated a tremendous potential to serve as a biomarker for neurologic and psychiatric disease [\citeonline{vosoughi2021schizophrenia, chockanathan2019automated, 54_abidin2018alteration, 56_dsouza2019classification}]. Currently, most of the diagnosis in brain-related disorders relies on clinical symptom evaluations, such as based on neuropsychological testing. However, there is a need for more objective biomarkers. To this end, more recently, studies have investigated, if such diagnostic information can be extracted non-invasively from brain activity data. Despite promising results, there is scope for improvement, specifically with regards to using more meaningful connectivity analysis approaches  [\citeonline{dadi2019benchmarking}].

Here, biomarkers from brain imaging using resting-state functional MRI (rs-fMRI) can be derived using Multi-Voxel Pattern Analysis (MVPA) techniques [\citeonline{dsouza2019multivoxel}]. MVPA is a machine-learning framework that extracts differences in patterns of brain connectivity for discriminating between connectivity profiles of individuals with neurological disorders and healthy individuals. Results of such studies demonstrate that clinically meaningful information can be learned from fMRI data. However, the simple conventional approach of using cross-correlation between time-series does not measure \textit{directed} connectivity. Hence, more relevant information in the fMRI data may be captured by more sophisticated connectivity measures which retain information about the direction of interdependence between time-series.

Various methods have been proposed to obtain such directional relationships in multivariate time-series data, including transfer entropy [\citeonline{schreiber2000measuring}] and mutual information [\citeonline{kraskov2004estimating}]. However, in high-dimensional systems, estimating the system’s underlying density function becomes computationally expensive [\citeonline{mozaffari2019online_ieee}]. Under the Gaussian assumption, transfer entropy is equivalent to Granger causality [\citeonline{barnett2009granger}]. However, the computation of multivariate Granger causality for short time series in large-scale problems is challenging  [\citeonline{vosoughi2021eusipco,dsouza2021large, vosoughi2021_LSAGC,vosoughi2021_lsNonlinear,vosoughi2021schizophrenia}]. Addressing this challenge, we recently proposed large-scale Extended Granger Causality (lsXGC), which is a method that combines the advantages of dimensionality reduction with the augmentation of a conditional source time-series adopted from the original space. The lsXGC method uses predictive time-series modeling for estimating directed causal relationships among fMRI time-series  [\citeonline{vosoughi2021schizophrenia}], which makes it a suitable method to quantify the uncertainty associated with the stochastic process. This work extends our previous investigation on lsXGC and evaluates its ability to accurately recovering directional information from synthetic rs-fMRI data with known ground truth (GT). Specifically, we compare the performance of lsXGC to conventional multivariate Granger causality (GC) [\citeonline{granger1988some}], mutual information (MI) [\citeonline{kraskov2004estimating}], and transfer entropy (TE) [\citeonline{schreiber2000measuring}].  Based on extensive simulations, we investigate the performance of lsXGC at correctly unveiling the underlying dynamics of the synthetic fMRI dataset, even at a significantly lower computational expense when compared to competing methods. Our results qualify lsXGC as a promising candidate for serving as a potential biomarker for brain disease in clinical fMRI studies. \newline
This work is embedded in our group’s endeavor to expedite artificial intelligence in biomedical imaging by means of advanced pattern recognition and machine learning methods for computational radiology and radiomics, e.g., [\citeonline{12_wismuller2006exploratory,
13_wismuller1998neural,
14_wismuller2002deformable,
15_behrends2003segmentation,
16_wismuller1997neural,
17_bunte2010exploratory,
18_wismuller1998deformable,
19_wismuller2009exploration,
20_wismuller2009method,
22_huber2010classification,
23_wismuller2009exploration,
24_bunte2011neighbor,
25_meyer2004model,
wismuller2001exploration,
saalbach2005hyperbolic,
wismuller2009computational,
26_wismuller2009computational,
leinsinger2003volumetric,
27_meyer2003topographic,
wismuller2009exploration,
28_meyer2009small,
29_wismueller2010model,
meyer2007unsupervised,
wismuller2000neural,
meyer2007analysis,
32_wismueller2008human,
wismuller2000hierarchical,
wismuller2015method,
33_huber2012texture,
37_otto2003model,
38_varini2004breast,
48_meyer2004computer,
40_meyer2004stability,
wismuller1998hierarchical,
41_meyer2008computer,
wismuller2013introducing,
45_bhole20143d,
46_nagarajan2013computer,
wismuller2001automatic,
wismueller1999adaptive,
meyer2005computer,
49_nagarajan2014computer,
pester2013exploring,
50_nagarajan2014classification,
yang2014improving,
wismuller2014pair,
wang2014investigating,
51_wismuller2014framework,
meyer2004local,
schmidt2014impact,
nagarajan2015integrating,
wismuller2015nonlinear,
nagarajan2015characterizing,
abidin2015volumetric,
wismuller2016mutual,
abidin2016investigating,
52_schmidt2016multivariate,
abidin2017classification,
abidin2017using,
61_dsouza2017exploring,
55_abidin2018deep,
chockanathan2018resilient,
dsouza2018mutual,
abidin2019investigating,
abidin2020detecting,
wismuller2020prospective,
dsouza2020large,
vosoughi2020large,
vosoughi2021marijuana, 
vosoughi2022lsKGC, 
wismuller2022lsNGC, 
wismuller2022lsAGC, 
Vosoughi2022_ICASSP
}].

\section{METHODS}\label{sec:methods}
\subsection{Large-scale Extended Granger Causality (lsXGC)}

Large-scale Extended Granger Causality (lsXGC) has been developed based on 1) the principle of original Granger
causality, which quantifies the causal influence of time-series $\bold{x_s}$ on time-series $\bold{x_t}$ by quantifying the amount of improvement in the prediction of $\bold{x_t}$ in presence of $\bold{x_s}$. 2) the idea of dimension reduction, which resolves the problem of the tackling a under-determined system, which is frequently faced in fMRI analysis, since the number of acquired temporal samples usually is not sufficient for estimating the model parameters [\citeonline{vosoughi2020large,dsouza2020large}].

Consider the ensemble of time-series $\mathcal{X}\in \mathbb{R}^{N\times T}$, where $N$ is the number of time-series (Regions Of Interest – ROIs) and $T$ the number of temporal samples. Let $\mathcal{X} = (\bold{x_1}, \bold{x_2}, \dots, \bold{x_N})^{\mathsf{T}}$ be the whole multidimensional system and $x_i \in \mathbb{R}^{1\times T}$ a single time-series with $i = 1, 2, \dots,N$, where $\bold{x_i} = (x_i(1), x_i(2), \dots, x_i(T))$. In order to overcome the under-determined problem, first $\mathcal{X}$ will be decomposed into its first $p$ high-variance principal components
$\mathcal{Z} \in \mathbb{R}^{p\times T}$ using Principal Component Analysis (PCA), \textit{i.e.},

\begin{equation}
\mathcal{Z}=W\mathcal{X},    
\end{equation}

where $W\in \mathbb{R}^{p\times N}$ represents the PCA coefficient matrix. Subsequently, the dimension-reduced time-series ensemble $\mathcal{Z}$ is augmented by one original time-series $\bold{x_s}$ yielding a dimension-reduced augmented time-series ensemble $\mathcal{Y}\in \mathbb{R}^{(p+1)\times T}$ for estimating the influence of $\bold{x_s}$ on all other time-series.

Following this, we locally predict $\mathcal{X}$ at each time sample $t$, \textit{i.e.} $\mathcal{X}(t)\in \mathbb{R}^{N\times 1}$ by calculating an estimate $\hat{\mathcal{X}}_{\bold{x_s}}(t)$. To this end, we fit an affine model based on a vector of $m$ vector of m time samples of $\mathcal{Y}(\tau)\in \mathbb{R}^{(p+1)\times 1}$($\tau=t-1, t-2, \dots, t-m$), which is $\bold{y}(t)\in \mathbb{R}^{m.(p+1)\times 1}$, and a parameter matrix $\mathcal{A}\in \mathbb{R}^{N\times m.(p+1)}$ and a constant bias vector $\bold{b}\in \mathbb{R}^{N\times 1}$, 
\begin{equation}
    \hat{\mathcal{X}}_{\bold{x_s}}(t)=\mathcal{A}\bold{y}(t)+\bold{b},~~ t=m+1, m+2, \dots, T.
\end{equation}

Now $\hat{\mathcal{X}}_{\setminus {\bold{x_s}}}(t)$, which is the prediction of $\mathcal{X}(t)$ without the information of $\bold{x_s}$, will be estimated. The estimation processes is identical to the previous one, with the only difference being that we have to remove the augmented time-series $\bold{x_s}$ and its corresponding column in the PCA coefficient matrix $W$.

The computation of a lsXGC index is based on comparing the variance of the prediction errors obtained with
and without consideration of $\bold{x_s}$. The lsXGC index $f_{\bold{x_s}\xrightarrow{}\bold{x_t}}$ , which indicates the influence of $\bold{x_s}$ on $\bold{x_t}$, can be calculated by the following equation:
\begin{equation}
    f_{\bold{x_s}\xrightarrow{}\bold{x_t}}=\log {\frac{\mathrm{var}(e_s)}{\mathrm{var}(e_{\setminus s})}},
\end{equation}
where $e_{\setminus s}$ is the error in predicting $\bold{x_t}$ when $\bold{x_s}$ was not considered, and $e_s$ is the error, when $\bold{x_s}$ was used. In this study, we set $p = 1$ and $m = 2$.

\section{Simulation on synthetic fMRI datasets}

\begin{figure}
    \centering
    \includegraphics[width=0.75\linewidth]{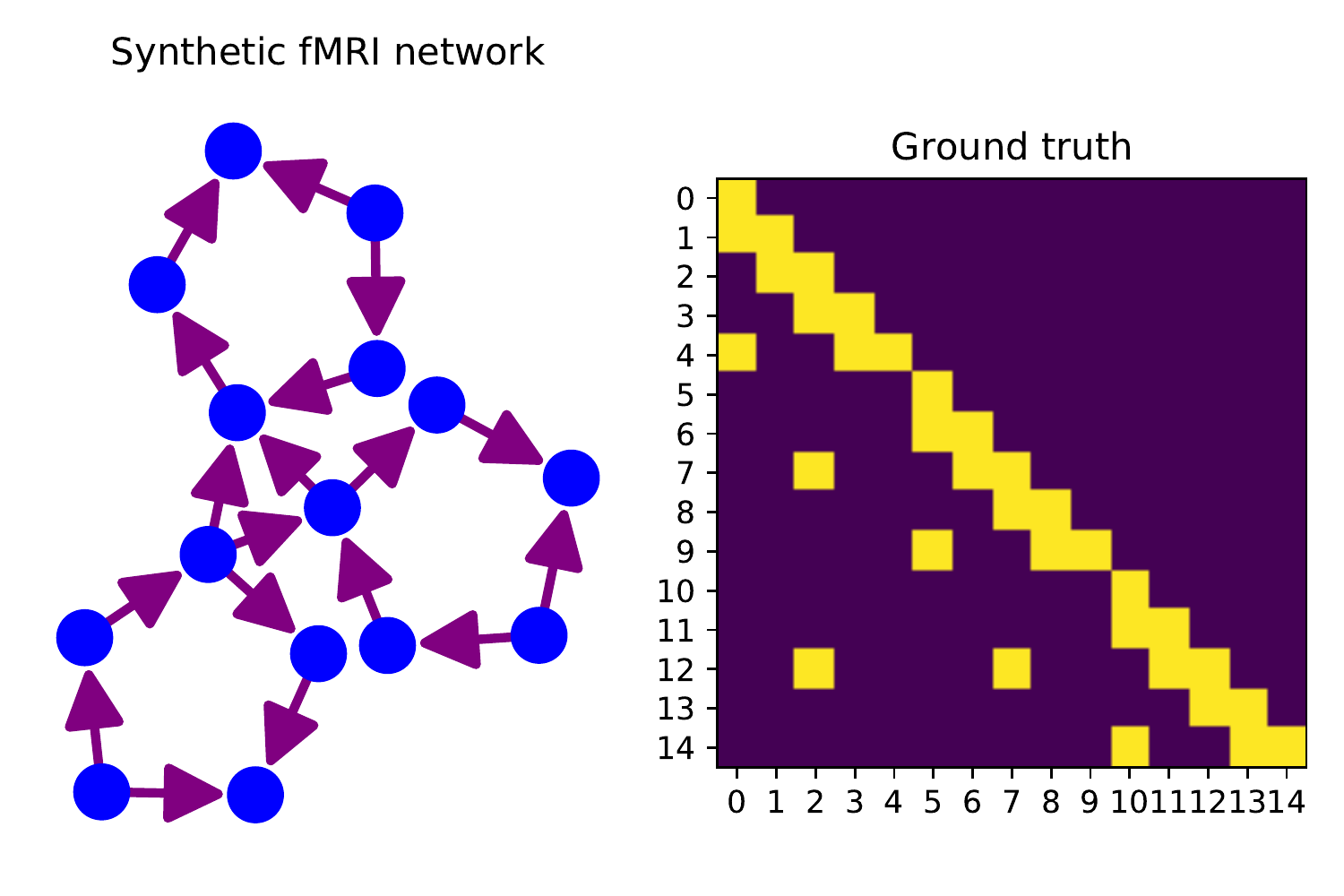}
     \caption{Netsim network 'sim3' adopted from [\citeonline{smith2011network}] and the ground truth of one of the networks is shown in this figure. The network simulations model hemodynamic response and noise, which resembles real-world resting-state fMRI data, with signal-to-noise ratio of 20 $dB$. The network has 15 nodes, and 50 different realizations of the network were used in our evaluations.} 
     \label{fig:graph_smith}
     
\end{figure}

Quantitative evaluation of lsXGC and competing algorithms was performed on semi-realistic data of rs-fMRI brain recordings [\citeonline{smith2011network, lowe2020amortized}], in which noise and the hemodynamic response were included in the data generation process. These data were adopted from Netsim [\citeonline{smith2011network}], and the network 'sim3' with $N=15$ was used in our simulations.  The hemodynamic responses were generated to model the time-delayed interactions among different brain regions, and, as a result, a smooth signal was generated that resembles recorded fMRI brain signals with an inherent signal-to-noise ratio (SNR) of 20 $dB$. The advantage of the simulated fMRI is that the ground truth (GT) for network connectivity is known, which is not available for real fMRI datasets. The graph and GT are shown in Fig.~\ref{fig:graph_smith}. Hence, the test results can be used to quantitatively compare different algorithms in the presence of noise hemodynamic response condition, which is not possible for real-world datasets. In line with the literature, we use the Area Under the Receiver Operating Characteristic (AUROC) as a comparison metric. Here, the task is to infer the underlying connectivity between 15 brain regions across 50 network simulation instances.

\begin{figure}
    \centering
    \includegraphics[width=0.65\linewidth]{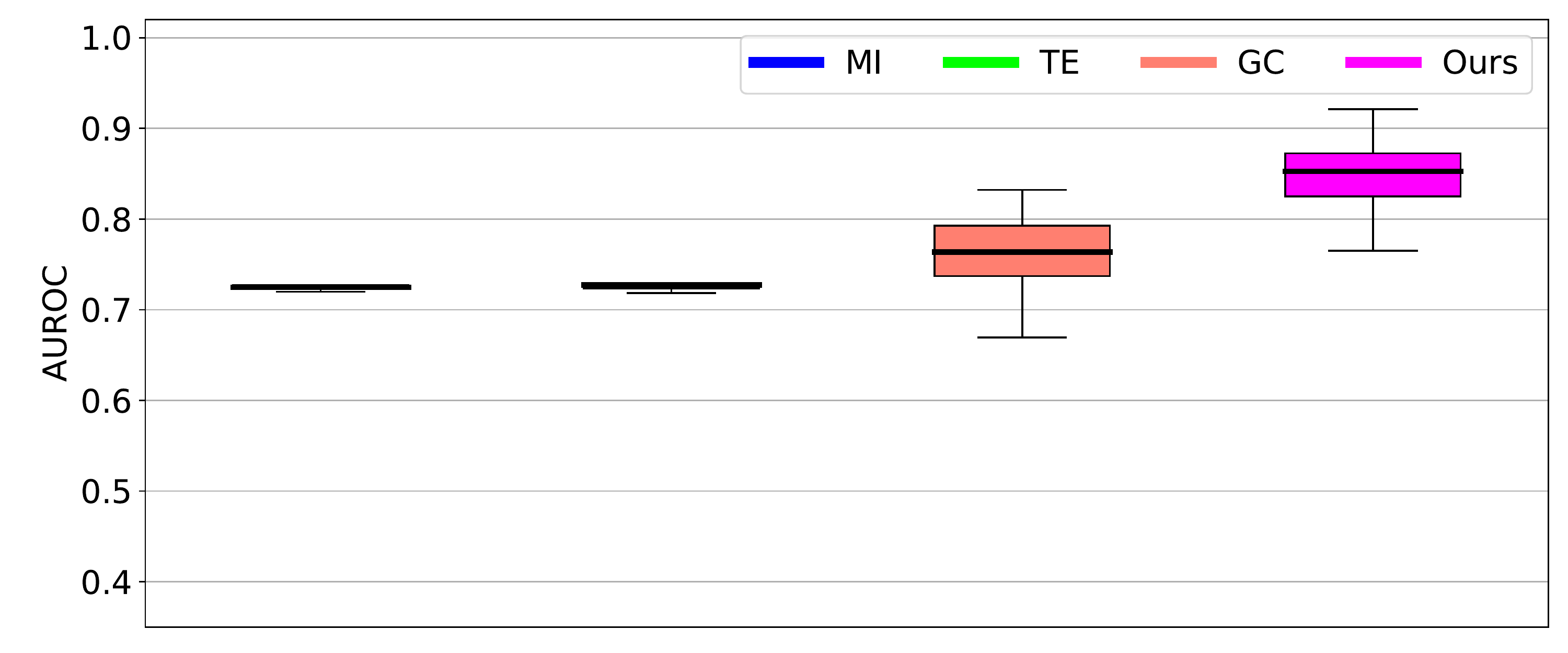}
     \caption{Performance of various algorithms, namely mutual information (MI) [\citeonline{kraskov2004estimating}], transfer entropy (TE) [\citeonline{schreiber2000measuring}], multivariate Granger causality (GC) [\citeonline{granger1988some}], and our proposed method (lsXGC, [\citeonline{vosoughi2021schizophrenia}]) on the Netsim dataset [\citeonline{smith2011network}]. Areas under the Receiver Operating Characteristics (AUROC) are shown in the figure, where each column represents a different algorithm (see titles). Boxplots represent [Q1, Q3]=[0.25, 0.75] quartiles and median, and the cap lines represent [minimum, maximum]=[Q1-1.5$\times$(Q3-Q1), Q3+1.5$\times$(Q3-Q1)]. As can be seen, the proposed algorithm (lsXGC) significantly outperforms other methods from the literature, suggesting robustness of the method against noise and hemodynamic response effects encountered in fMRI data. Wilcoxon test $p$-values of the proposed lsXGC method as compared to MI, TE, and GC are less than $\{<\!10^{-9}, <\!10^{-9}, <\!10^{-8}\}$, correspondingly.} 
     \label{fig:smith_AUC}
\end{figure}

Results are shown in Fig.~\ref{fig:smith_AUC} and Table.~\ref{tab:sim_time_smith}. In Fig.~\ref{fig:smith_AUC}, the AUROC of the proposed lsXGC algorithm significantly outperforms the competing methods, affirming its ability at accurately detecting the underlying directed network connectivity structure in simulated fMRI signals under conditions of noise and smoothing based on hemodynamic response, resembling the real-world situation. Simulation time and diagnostic accuracy results are listed in Table.~\ref{tab:sim_time_smith}.

\begin{SCtable}[]
\centering
\caption{Simulation time and performance on Netsim dataset [\citeonline{smith2011network}] in identical conditions, except TE and MI, which were performed on Nvidia GeForce 1080-Ti GPU, with the rest based on CPU operations. All simulations were performed in Python 3.8.}\label{tab:sim_time_smith}
\begin{tabular}{|c|c|c|}
\hline
      & Time (seconds) & AUROC                 \\ \hline
MI    & 4753             & 0.728 $\pm$ 0.010  \\ \hline
TE    & 2806             & 0.727 $\pm$ 0.006  \\ \hline
GC    & 9.7             & 0.762 $\pm$ 0.038  \\ \hline
lsXGC & \textbf{3.4}             & \textbf{0.849 + 0.032}  \\ \hline
\end{tabular}
\end{SCtable}

\section{CONCLUSIONS}\label{sec:conclusions}

This paper investigates a novel method, large-scale Extended Granger Causality (lsXGC), for discovering relations in high-dimensional dynamical systems involving observational data. The lsXGC method addresses the curse of dimensionality in large-scale dynamic systems by combining dimensionality reduction with data augmentation, namely by augmenting data from the conditional source time-series in the original input space onto a dimensionally reduced state-space representation of the multidimensional time-series system, thus improving prediction quality in a Granger causality setting. Here, we focus on numerical results for learning meaningful representations from synthetic resting-state fMRI data, although there may be a wide scope of other application domains, ranging from neuroscience to climatology and finance. Our results suggest competitive performance of the proposed method in comparison to other widely used techniques from the literature, namely conventional Granger causality, transfer entropy, and mutual information. Most importantly, our method infers network connectivity at a significantly lower computational expense, while accurately identifying causal representations from multidimensional time-series data. We conclude that lsXGC exhibits important advantages over competing methods from the existing literature for inferring directed causal network connectivity in large-scale systems with limited observational data. Specifically, our results suggest that lsXGC may qualify as a promising candidate for serving as a potential biomarker for brain disease in clinical fMRI studies.

\renewcommand{\baselinestretch}{0.5}
\footnotesize

\acknowledgments 
 
This research was partially funded by the American College of Radiology (ACR) Innovation Award “AI-PROBE: A Novel Prospective Randomized Clinical Trial Approach for Investigating the Clinical Usefulness of Artificial Intelligence in Radiology” (PI: Axel Wism{\"u}ller) and an Ernest J. Del Monte Institute for Neuroscience Award from the Harry T. Mangurian Jr. Foundation (PI: Axel Wism{\"u}ller). This work was conducted as a Practice Quality Improvement (PQI) project related to American Board of Radiology (ABR) Maintenance of Certificate (MOC) for A.W.

\newpage
\bibliography{report} 
\bibliographystyle{spiebib} 

\end{document}